\documentstyle[prl,eqsecnum,aps]{revtex}
\begin{document}
\author{Jian Qi Shen \footnote{E-mail address: jqshen@coer.zju.edu.cn}}
\address{Zhejiang Institute of Modern Physics and Department of Physics,\\
Zhejiang University, Hangzhou 310027, People$^{,}$s Republic of China}
\date{\today}
\title{Gravitational Analogues, Geometric Effects and Gravitomagnetic Charge \footnote{This paper has been published in Gen. Rel. Gra. {\bf 34}(9), 1423-1435 (2002) and the present e-print form is the revised version of the published paper.}}
\maketitle

\begin{abstract}
This essay discusses some geometric effects associated with
gravitomagnetic fields and gravitomagnetic charge as well as the
gravity theory of the latter. Gravitomagnetic charge is the
duality of gravitoelectric charge ( mass ) and is therefore also
termed the {\it dual mass} which represents the topological
property of gravitation. The field equation of gravitomagnetic
matter is suggested and a static spherically symmetric solution of
this equation is offered. A possible explanation of the anomalous
acceleration acting on Pioneer spacecrafts are briefly proposed.

\end{abstract}

\pacs{PACS:}

\section{INTRODUCTION}

Considering the following gravitational analogues of
electromagnetic phenomena is of physical interest: (1) in
electrodynamics a charged particle is acted upon by the Lorentz
magnetic force and, in the similar fashion, a particle is also
acted upon by the gravitational Lorentz force in weak-gravity
theory \cite{Shen,Kleinert}. According to the principle of
equivalence, further analysis shows that in the non-inertial
rotating reference frame, this gravitational Lorentz force is just
the fictitious Coriolis force; (2) there exists Aharonov-Bohm
effect in electrodynamics\cite {Bohm}, accordingly, the so-called
gravitational Aharonov-Bohm effect, {\it i.e.}, the gravitational
analogue of Aharonov-Bohm effect also exists in the theory of
gravitation, which is now termed the Aharonov-Carmi effect\cite
{Aharonov,Overhauser,Werner}; (3) a particle with intrinsic spin
possesses a gravitomagnetic moment\cite{Hehl} of such magnitude
that it equals the spin of this particle. The interaction of
spinning gravitomagnetic moment with the gravitomagnetic field is
called spin-gravity coupling, which is similar to the interaction
between spinning magnetic moment and magnetic fields in
electrodynamics; (4) in the rotating reference frame, the rotating
frequency relative to the fixing frame may be considered the
effective gravitomagnetic field strength that is independent of
the Newtonian gravitational constant, $G$, in accordance with the
principle of equivalence. This, therefore, means that the nature
of spin-rotation coupling \cite{Mashhoon1,Mashhoon2} is the
interaction of spinning gravitomagnetic moment with
gravitomagnetic fields; (5) it is well known that geometric phase
reflecting the global and topological properties of evolution of
the quantum systems\cite{Berry,Simon} appears in systems whose
Hamiltonian is time-dependent or possessing evolution parameters.
Apparently, the geometric phase in the Aharonov-Bohm effect and
Aharonov-Carmi effect results from the presence of the evolution
parameter in the Hamiltonian. We suggested another geometric
phase\cite{Shen} that exists in the spin-rotation coupling system
where the rotating frequency of the rotating frame is
time-dependent. Investigation of this
geometric phase is believed to be a potential application to the Earth$%
^{^{,}}$s time-dependent rotating frequency ( frequency
fluctuations ), namely, by measuring the geometric phase
difference of spin polarized vertically down and up in the
neutron-gravity interferometry experiment, one may obtain the
information concerning the variation of the Earth$^{^{,}}$s
rotation.

For the present, it is possible to investigate quantum mechanics
in weak-gravitational fields\cite{Lamm,Alvarez}, with the
development of detecting and measuring technology, particularly
laser-interferometer technology, low-temperature technology,
electronic technology and so on. These investigations enable
physicists to test validity or universality of fundamental laws
and principles of general relativity. It should be noted that the
Aharonov-Carmi effect and the geometric phase factor in the
time-dependent spin-rotation coupling reflect the aspects of
geometric properties in gravity. Both of them are related to the
gravitomagnetic fields. In this essay, the author discusses
another geometric or topological aspect of the gravity, {\it
i.e.}, the gravitomagnetic charge that is the gravitational
analogue of magnetic monopole in electrodynamics.

In electrodynamics, electric charge is a Noether charge while its
dual charge ( magnetic charge ) is a topological charge, since the
latter relates to the singularity of non-analytical vector
potentials. Magnetic monopole \cite{Dirac} attracts attentions of
many physicists in various fields such as gauge field theory,
grand unified theory, particle physics and cosmology
\cite{Schwinger,Yang1,Yang2,Hooft,Tchra,Chak}. In the similar
fashion, it is also interesting to consider the gravitomagnetic
charge, which is the source of gravitomagnetic field just as mass
( gravitoelectric charge ) is the source of gravitoelectric field
( Newtonian gravitational field ). In this sense, gravitomagnetic
charge is also termed {\it dual mass}. It should be noted that the
concept of mass is of no significance for the gravitomagnetic
charge and it is therefore of interest to investigate the
relativistic dynamics and gravitational effects of this
topological {\it dual mass}.

The paper is organized as follows: The gravitational field
equation of {\it gravitomagnetic matter } is derived in Sec. II,
the form of field equation in the weak-field approximation is
given in Sec. III and the exact static spherically symmetric
solution is obtained in Sec. IV. In Sec. V, two related problems,
{\it i.e.}, the geometric phase factor possessed by a photon
propagating in the gravitomagnetic field, and a potential
interpretation, by means of the mechanism of gravitational
Meissner effect, of the anomalous acceleration acting on Pioneer
spacecrafts\cite{Anderson} are briefly proposed. In Sec. VI, the
author concludes with some remarks.

\section{GRAVITATIONAL FIELD EQUATION OF GRAVITOMAGNETIC MATTER}

In order to obtain the gravitational field equation of gravitomagnetic
charge, we should construct the dual Einstein$^{,}$s tensor. By using the
variational principle, we can obtain the following equation
\begin{equation}
\delta \int_{\Omega }\sqrt{-g}\tilde{R}{\rm d}\Omega =\int_{\Omega }\sqrt{-g}%
\tilde{G}_{\mu \nu }\delta g^{\mu \nu }{\rm d}\Omega  \eqnum{1}
\label{eq1}
\end{equation}
with
\begin{equation}
\tilde{R}=g^{\sigma \tau }\tilde{R}_{\sigma \tau },\quad
\tilde{R}_{\upsilon \gamma}=g^{\mu \delta }(\epsilon _{\mu
\upsilon }^{\quad \ \alpha \beta }R_{\gamma \delta \alpha \beta
}+\epsilon _{\gamma \delta }^{\quad \ \alpha \beta }R_{\mu
\upsilon \alpha \beta }) \eqnum{2}  \label{eq2}
\end{equation}
and
\begin{equation}
\tilde{G}_{\mu \nu }=\epsilon _{\mu }^{\quad \lambda \sigma \tau }R_{\nu
\lambda \sigma \tau }-\epsilon _{\nu }^{\quad \lambda \sigma \tau }R_{\mu
\lambda \sigma \tau },  \eqnum{3}  \label{eq3}
\end{equation}
where\footnote{When it comes to the description problem of fields
arising from the dual charges, the dual tensor $\tilde{G}_{\mu \nu
}$ in gravity theory is analogous to the dual field tensor
$\tilde{F}^{\mu \nu }=\frac{1}{2}\epsilon ^{\mu \nu \alpha \beta
}F_{\alpha \beta }$ in electrodynamics.} $\epsilon_{\mu }^{\quad
\lambda \sigma \tau }=g_{\mu \nu }\epsilon ^{\nu \lambda \sigma
\tau }$ with $\epsilon ^{\nu \lambda \sigma \tau }$
being the completely antisymmetric Levi-Civita tensor, and the dual scalar curvature $%
\tilde{R}$ is assumed to be the Lagrangian density of the
interaction of metric fields with gravitomagnetic matter. Since
the dual Einstein$^{,}$s tensor, $\tilde{G}_{\mu \nu }$, is an
antisymmetric tensor, we construct the following antisymmetric
tensor for the Fermi field
\begin{equation}
K_{\mu \nu }=i\bar{\psi}(\gamma _{\mu }\partial _{\nu }-\gamma _{\upsilon
}\partial _{\mu })\psi ,\quad H_{\mu \nu }=\epsilon _{\mu \nu }^{\quad
\alpha \beta }K_{\alpha \beta }  \eqnum{4}  \label{eq4}
\end{equation}
and regard them as the source tensors in the field equation of
gravitomagnetic charge, where $\gamma _{\mu }^{^{,}}$s are general
Dirac matrices with respect to $x^{\mu }$ and satisfy $\gamma
_{\mu }\gamma _{v}+\gamma _{\nu }\gamma _{\mu }=2g_{\mu \nu }.$
Then the field equation of gravitational field produced by the
gravitomagnetic charge may be given as follows\footnote{One of the
advantages of this field equation is that it does not introduce
extra tensor potentials when allowing for gravitomagnetic monopole
densities and currents. This fact is in analogy with that in
electrodynamics, where the equation $\partial _{\nu
}\tilde{F}^{\mu \nu }=J_{\rm \bf M}^{\mu}$ governs the motion of
electromagnetic fields produced by magnetic monopoles.}
\begin{equation}
\tilde{G}_{\mu \nu }=\kappa _{1}K_{\mu \nu }+\kappa _{2}H_{\mu \nu }
\eqnum{5}  \label{eq5}
\end{equation}
with $\kappa _{1},\kappa _{2}$ being the coupling coefficients
between {\it gravitomagnetic matter } and gravity. It should be
noted that $\tilde{G}_{\mu \nu }\equiv 0$ in the absence of
gravitomagnetic matter since no singularities associated with
topological charge exist in the metric functions and therefore
Ricci identity still holds. However, once the metric functions
possess non-analytic properties in the presence of gravitomagnetic
matter ( should such exist ), $\tilde{G}_{\mu \nu }$, is no longer
vanishing due to the violation of Ricci identity. Additionally,
further investigation shows that the {\sl cosmological term }of
Fermi field in Eq. (\ref{eq5}) can
be written as the linear combination of the antisymmetric tensors $i\bar{\psi%
}(\gamma _{\mu }\gamma _{\nu }-\gamma _{\upsilon }\gamma _{\mu })\psi $ and $%
i\epsilon _{\mu \nu }^{\quad \alpha \beta }\bar{\psi}(\gamma _{\alpha
}\gamma _{\beta }-\gamma _{\beta }\gamma _{\alpha })\psi $.

It is believed that there would exist formation (and creation)
mechanism of gravitomagnetic charge in the gravitational
interaction, just as some prevalent theories\cite{Hooft} provide
the theoretical mechanism of existence of magnetic monopole in
various gauge interactions. Magnetic monopole in electrodynamics
and gauge field theory has been discussed and sought after for
decades, and the existence of the $^{,}$t Hooft-Polyakov monopole
solutions \cite{Hooft,Polyakov,Polyakov2} has spurred new interest
of both theorists and experimentalists. Similar to magnetic
monopole, gravitomagnetic charge is believed to give rise to such
situations. If it is indeed present in universe, it will also lead
to significant consequences in astrophysics and cosmology. We
emphasize that although it is the classical solution to the field
equation as discussed above, this kind of topological
gravitomagnetic monopoles may arise not as fundamental entities in
gravity theory.

\section{LOW-MOTION WEAK-FIELD APPROXIMATION}

In what follows the low-motion weak-field approximation is applied
to the following gravitational field equation of {\it
gravitomagnetic matter }
\begin{equation}
\tilde{G}^{\mu \nu }=S^{\mu \nu }  \eqnum{6}  \label{eq6}
\end{equation}
with the source tensor $S^{\mu \nu }=\kappa _{1}K^{\mu \nu }+\kappa
_{2}H^{\mu \nu }$. First we consider $\tilde{G}^{01}=\epsilon ^{0\alpha
\beta \gamma }R_{\quad \alpha \beta \gamma }^{1}-\epsilon ^{1\alpha \beta
\gamma }R_{\quad \alpha \beta \gamma }^{0}$ by the linear approximation. The
following expressions may be given
\begin{eqnarray}
-\epsilon ^{1\alpha \beta \gamma }R_{\quad \alpha \beta \gamma }^{0} &\simeq
&2(R_{0302}+R_{0230}),  \nonumber \\
2R_{0302} &\simeq &\frac{\partial ^{2}g_{02}}{\partial x^{3}\partial x^{0}}+%
\frac{\partial ^{2}g_{30}}{\partial x^{0}\partial x^{2}}-\frac{\partial
^{2}g_{00}}{\partial x^{3}\partial x^{2}}-\frac{\partial ^{2}g_{32}}{%
\partial x^{0}\partial x^{0}},  \nonumber \\
2R_{0230} &\simeq &\frac{\partial ^{2}g_{00}}{\partial x^{2}\partial x^{3}}+%
\frac{\partial ^{2}g_{23}}{\partial x^{0}\partial x^{0}}-\frac{\partial
^{2}g_{03}}{\partial x^{2}\partial x^{0}}-\frac{\partial ^{2}g_{02}}{%
\partial x^{0}\partial x^{3}},  \eqnum{7}  \label{eq7}
\end{eqnarray}
where the nonlinear terms (the products of two Christoffel
symbols) are ignored and use is made of $\epsilon ^{1023}=\epsilon
^{1302}=\epsilon ^{1230}\simeq -1$ and $R_{\quad 023}^{0}\simeq
R_{0023}=0,$ $R_{\quad 302}^{0}\simeq R_{0302},$ $R_{\quad
230}^{0}\simeq R_{0230}.$ If metric functions $g_{\mu \nu }$ are
analytic, then $2(R_{0302}+R_{0230})$ is therefore vanishing. But
once gravitomagnetic charge is present in spacetime and thus the
metric functions possess the singularities, this result does
not hold. Taking the gravitomagnetic vector potential ${\bf {g}}%
=(g^{01},g^{02},g^{03}),$ $-{\bf {g}}=(g_{01},g_{02},g_{03}),$ one
can arrive at
\begin{equation}
-\epsilon ^{1\alpha \beta \gamma }R_{\quad \alpha \beta \gamma }^{0}=-\frac{%
\partial }{\partial x^{0}}(\nabla \times {\bf {g}})_{1}-\nabla \times (\nabla
g^{00}-\frac{\partial {\bf {g}}}{\partial x^{0}})_{1}.  \eqnum{8}
\label{eq8}
\end{equation}
When we utilize the linear approximation for the field equation of
{\it dual matter} ( gravitomagnetic matter ), we are concerned
only with the space-time derivatives of gravitational potentials
$g^{\mu
}=(\frac{g^{00}-1}{2},g^{01},g^{02},g^{03})$ rather than that of $g_{ii}$ and $g_{ij}$ with $%
i,j=1,2,3,$ since the latter is either the analytic functions or the small
terms. This, therefore, implies that the contribution of $\epsilon ^{0\alpha
\beta \gamma }R_{\quad \alpha \beta \gamma }^{1}$ to $\tilde{G}^{01}$
vanishes. Eq. (\ref{eq8}) is readily generalized to the three-dimensional
case, and the combination of Eq. (\ref{eq6}) and Eq. (\ref{eq8}) yields
\begin{equation}
\nabla \times (\nabla g^{00}-\frac{\partial }{\partial x^{0}}{\bf {g}})=-\frac{%
\partial }{\partial x^{0}}(\nabla \times {\bf {g}})+\vec{S},  \eqnum{9}
\label{eq9}
\end{equation}
where $\vec{S}$ is defined to be $S^{i0}$ $(i=1,2,3).$ It is apparently seen
that Eq. (\ref{eq9}) is exactly analogous to the Faraday$^{^{,}}$s law of
electromagnetic induction in the presence of current density of magnetic
monopole in electrodynamics. This, therefore, implies that Eq. (\ref{eq6})
is indeed the field equation of gravitation of gravitomagnetic matter.

It is also of interest to discuss the motion of gravitomagnetic monopole in
curved spacetime. Although Ricci identity is violated due to the
non-analytic properties caused by the gravitomagnetic charge, Bianchi
identity still holds in the presence of gravitomagnetic charge. It follows
that the covariant divergence of $\tilde{G}^{\mu \nu }$ vanishes, namely,
\begin{equation}
\tilde{G}_{\quad ;\nu }^{\mu \nu }=0.  \eqnum{10}  \label{eq12}
\end{equation}
Then in terms of the following field equation
\begin{equation}
\tilde{G}^{\mu \nu }=S^{\mu \nu }  \eqnum{11}  \label{eq13}
\end{equation}
with the antisymmetric source tensor of gravitomagnetic matter $S^{\mu \nu }$
being $\kappa _{1}K^{\mu \nu }+\kappa _{2}H^{\mu \nu }$, one can arrive at
\begin{equation}
S_{\quad ;\nu }^{\mu \nu }=0  \eqnum{12}  \label{eq14}
\end{equation}
which may be regarded as the equation of motion of gravitomagnetic charge in
the curved spacetime. It is useful to obtain the low-motion and
weak-field-approximation form of Eq. (\ref{eq14}), which enables us to
guarantee that Eq. (\ref{eq14}) is the equation of motion of gravitomagnetic
monopole indeed.

The general Dirac matrices in the weak-field approximation may be obtained
via the relations $\gamma _{\mu }\gamma _{v}+\gamma _{\nu }\gamma _{\mu
}=2g_{\mu \nu }$ and the results are given by
\begin{equation}
\gamma ^{0}=(1+g^{0})\beta ,\quad \gamma ^{i}=g^{i}\beta +\gamma _{M}^{i},
\eqnum{13}  \label{eq15}
\end{equation}
where $i=1,2,3;$ $\beta =\gamma _{M}^{0}.$ $\gamma _{M}^{0}$ and $\gamma
_{M}^{i}$ are the constant Dirac matrices in the flat Minkowski spacetime.
Note that the gravitoelectric potential is defined to be $g^{0}=\frac{%
g^{00}-1}{2},$ and gravitomagnetic vector potentials are $g^{i}=g^{0i}\quad
(i=1,2,3).$ In the framework of dynamics of point-like particle, the source
tensor is therefore rewritten as
\begin{equation}
S^{\mu \nu }=\rho \left[ \kappa _{1}(g^{\mu }U^{\nu }-g^{\nu }U^{\mu
})+\kappa _{2}\epsilon ^{\mu \nu \alpha \beta }(g_{\alpha }U_{\beta
}-g_{\beta }U_{\alpha })\right] ,  \eqnum{14}  \label{eq16}
\end{equation}
where $\rho $ denotes the density of gravitomagnetic matter. It follows from
Eq. (\ref{eq14}) and Eq. (\ref{eq16}) that there exists the gravitational
Lorentz force density in the expression for the force acting on the
gravitomagnetic charge, namely, the equation of motion of gravitomagnetic
charge is of the form
\begin{equation}
\frac{\partial }{\partial x^{0}}{\bf{v}}=[\nabla \times {\bf
{g}}-{\bf{v}}\times (\nabla g^{0}-\frac{\partial }{\partial
x^{0}}{\bf {g}})],  \eqnum{15} \label{eq17}
\end{equation}
where some small terms are ignored and the relation, $\kappa
_{1}g^{0}=2\kappa _{2}$, between the coupling coefficients,
$\kappa _{1}$ and $\kappa _{2}$ is assumed ( one may be referred
to the Appendix for the consideration why this assumption is
reasonable ). Note, however, that the relation of the two coupling
parameters suggested above holds only when weak-field
approximation is employed ( see Appendix for more detailed
analysis ). This connection between $\kappa _{1}$ and $\kappa
_{2}$ gives us a helpful insight into the generally covariant
relation between them.

In view of what has been discussed, it can be seen that, in the weak
gravitational field, the gravitomagnetic charge behaves like the magnetic
charge. This, therefore, implies that gravitomagnetic charge proposed above
is the gravitational analogue of magnetic charge in electrodynamics.

\section{EXACT SOLUTION OF STATIC SPHERICALLY SYMMETRIC GRAVITOMAGNETIC FIELD%
}
A static spherically symmetric solution is exactly obtained by
supposing that when a point-like gravitomagnetic charge is fixed
at the origin of the spherical coordinate system, the exterior
spacetime interval is of the form
\begin{equation}
{\rm d}s^{2}=({\rm d}x^{0})^{2}-{\rm d}r^{2}-r^{2}({\rm d}\theta
^{2}+\sin ^{2}\theta {\rm d}\varphi ^{2})+2g_{0\varphi }(\theta
){\rm d}x^{0}{\rm d}\varphi , \eqnum{16} \label{eq18}
\end{equation}
where the gravitomagnetic potential $g_{0\varphi }$ is assumed to be the
function with respect only to $\theta $.

Thus we obtain all the Christoffel symbols with non-vanishing
values as follows:
\begin{equation}
\Gamma _{0,\varphi \theta }=\Gamma _{0,\theta \varphi }=\frac{1}{2}\frac{%
\partial g_{0\varphi }}{\partial \theta },\quad \Gamma _{\varphi ,0\theta
}=\Gamma _{\varphi ,\theta 0}=\frac{1}{2}\frac{\partial g_{0\varphi }}{%
\partial \theta },\quad \Gamma _{\theta ,0\varphi }=\Gamma _{\theta ,\varphi
0}=-\frac{1}{2}\frac{\partial g_{0\varphi }}{\partial \theta }.  \eqnum{17}
\label{eq19}
\end{equation}
\ Since the field equation of gravitomagnetic matter is the antisymmetric
equation, we might as well take into account a simple case of the following
equation
\begin{equation}
\epsilon ^{0\alpha \beta \gamma }R_{\quad \alpha \beta \gamma }^{0}=M\delta
(x^{i})  \eqnum{18}  \label{eq20}
\end{equation}
with $M$ being the parameter associated with the coupling parameters and
gravitomagnetic charge. It is therefore apparent that Eq. (\ref{eq20})
agrees with Eq. (\ref{eq13}). Hence, the solution of the former equation
also satisfies the latter. For the reason of the completely antisymmetric
property of the Levi-Civita tensor, the contravariant indices $\alpha ,\beta
,\gamma $ should be respectively taken to be $x,y,z$ of three-dimensional
space coordinate, namely, we have
\begin{equation}
\epsilon ^{0\alpha \beta \gamma }R_{\quad \alpha \beta \gamma
}^{0}=2\epsilon ^{0xyz}(R_{\quad xyz}^{0}+R_{\quad zxy}^{0}+R_{\quad
yzx}^{0}).  \eqnum{19}  \label{eq21}
\end{equation}
There exist the products of two Christoffel symbols, {\it i.e.},
$g^{\sigma \tau }(\Gamma _{\tau ,\alpha \gamma }\Gamma _{\lambda
,\sigma \beta }-\Gamma _{\tau ,\alpha \beta }\Gamma _{\lambda
,\sigma \gamma })$ in the definition of the Riemann curvature,
$R_{\lambda \alpha \beta \gamma }$. Apparently, the products of
two Christoffel symbols (the nonlinear terms of field equation)
contain the total indices of three-dimensional space coordinate
(namely, these indices are taken the permutations of $\ r,\theta
,\varphi $ ) and therefore vanish, in the light of the fact that
the Christoffel symbol with index $r$ is vanishing in terms of Eq.
(\ref{eq19}).

In view of the above discussion, one can conclude that Eq. (\ref{eq20}) can
be exactly reduced to a linear equation. It is easily verified that $%
R_{\lambda \alpha \beta \gamma }$ $(\lambda =r,\theta ,\varphi )$ vanishes
with the help of the linear expression for $R_{\lambda \alpha \beta \gamma }$
given by $R_{\lambda \alpha \beta \gamma }=\frac{1}{2}(\frac{\partial
^{2}g_{\lambda \gamma }}{\partial x^{\alpha }\partial x^{\beta }}+\frac{%
\partial ^{2}g_{\alpha \beta }}{\partial x^{\lambda }\partial x^{\gamma }}-%
\frac{\partial ^{2}g_{\lambda \beta }}{\partial x^{\alpha }\partial
x^{\gamma }}-\frac{\partial ^{2}g_{\alpha \gamma }}{\partial x^{\lambda
}\partial x^{\beta }})$ and the linear element expressed by Eq. (\ref{eq18}%
). We thus obtain that $R_{\quad \alpha \beta \gamma }^{0}=g^{00}R_{0\alpha
\beta \gamma }.$ By the aid of the following expression
\begin{equation}
R_{0\alpha \beta \gamma }=\frac{1}{2}\frac{\partial }{\partial x^{\alpha }}(%
\frac{\partial g_{0\gamma }}{\partial x^{\beta }}-\frac{\partial g_{0\beta }%
}{\partial x^{\gamma }}),  \eqnum{20}  \label{eq22}
\end{equation}
one can arrive at
\begin{equation}
\epsilon ^{0\alpha \beta \gamma }R_{\quad \alpha \beta \gamma }^{0}=-\frac{%
g^{00}}{\sqrt{-g}}\nabla \cdot (\nabla \times {\bf {g}}),
\eqnum{21} \label{eq23}
\end{equation}
where the gravitomagnetic vector potentials, ${\bf {g}}$, are defined to be $%
{\bf {g}}=(-g_{0x},-g_{0y},-g_{0z})$. Substitution of Eq.
(\ref{eq23}) into Eq. (\ref{eq20}) yields
\begin{equation}
\nabla \cdot (\nabla \times {\bf
{g}})=-\frac{\sqrt{-g}}{g^{00}}M\delta (x^{i}).  \eqnum{22}
\label{eq24}
\end{equation}
Note that Eq. (\ref{eq24}) is the exact static gravitational field equation
of gravitomagnetic matter derived from Eq. (\ref{eq13}), where use is made
of the expression (\ref{eq18}) for the spacetime interval.

It follows from Eq. (\ref{eq24}) that the static spherically symmetric
solution is given as follows
\begin{equation}
2g_{0\varphi }{\rm d}x^{0}{\rm d}\varphi =\mp \frac{2c}{4\pi
}\cdot \frac{1\pm \cos \theta }{r\sin \theta }\cdot r\sin \theta
{\rm d}x^{0}{\rm d}\varphi ,  \eqnum{23} \label{eq25}
\end{equation}
where $c$ is defined to be determined by the metric functions of
the origin of the spherical coordinate system, {\it i.e.},
$c=-M(\frac{\sqrt{-g}}{g^{00}})_{\rm origin}$. Further calculation
yields
\begin{equation}
\left( g^{\mu \nu }\right) =\left(
\begin{array}{cccc}
\frac{r^{2}\sin ^{2}\theta }{r^{2}\sin ^{2}\theta +g_{0\varphi }^{2}} & 0 & 0
& \frac{g_{0\varphi }}{r^{2}\sin ^{2}\theta +g_{0\varphi }^{2}} \\
0 & -1 & 0 & 0 \\
0 & 0 & -r^{-2} & 0 \\
\frac{g_{0\varphi }}{r^{2}\sin ^{2}\theta +g_{0\varphi }^{2}} & 0 & 0 &
\frac{-1}{r^{2}\sin ^{2}\theta +g_{0\varphi }^{2}}
\end{array}
\right)  \eqnum{24}  \label{eq26}
\end{equation}
which is the inverse matrix of the metric $\left( g_{\mu \nu }\right) $ and
we thus obtain the contravariant metric $g^{\mu \nu }$. The gravitomagnetic
field strength takes the form
\begin{equation}
B_{\rm g}^{i}=\frac{c}{4\pi }\frac{x^{i}}{r^{3}},  \eqnum{25}
\label{eq200}
\end{equation}
provided that use is made of ${\bf{B}}_{\rm g}=\nabla \times
{\bf{A}}$ with ${\bf{A}}=(0,0,\mp \frac{2c}{4\pi }\frac{1\pm \cos
\theta }{r\sin \theta })$ expressed in spherical coordinate
system.

From what has been discussed above, similar to the magnetic charge in
electrodynamics, gravitomagnetic charge is a kind of topological charge that
is the duality of mass of matter. In this sense, gravitomagnetic charge is
also called dual mass. From the point of view of the classical field
equation, matter may be classified into two categories: gravitomagnetic
matter and gravitoelectric matter, of which the field equation of the latter
is Einstein$^{,}$s equation of general relativity. Due to their different
gravitational features, the concept of mass is of no significance for the
gravitomagnetic matter.

\section{TWO RELATED PROBLEMS}

(1) It is worthwhile to take into account the motion of photon in
gravitomagnetic fields. Consider a weak gravitomagnetic field
where the adiabatic approximation can be applicable to the motion
of a photon, a
conclusion can be drawn that the eigenvalue of the helicity $\frac{{\bf{k}}(t)%
}{k}\cdot {\bf{J}}$ of the photon is conserved in motion and the
helicity operator $\frac{{\bf{k}}(t)}{k}\cdot {\bf{J}}$ is an
invariant in terms of the invariant equation ( {\it i.e.}, the
Liouville-Von Neumann equation ) in Lewis-Riesenfeld
theory\cite{Lewis}
\begin{equation}
\frac{\partial I(t)}{\partial t}+\frac{1}{i}[I(t),H(t)]=0,  \eqnum{26}
\label{eq251}
\end{equation}
where the invariant $I(t)=\frac{{\bf{k}}(t)}{k}\cdot {\bf{J}}.$
From Eq. (\ref {eq251}), simple calculation yields
\begin{equation}
H(t)=\frac{{\bf{k}}\times \frac{\rm d}{{\rm
d}t}{\bf{k}}}{k^{2}}\cdot {\bf{J}}, \eqnum{27} \label{eq261}
\end{equation}
which is considered an effective Hamiltonian governing the motion of photon
in gravitomagnetic fields. Hence, the equation of motion of a photon in
gravitomagnetic fields is written
\begin{equation}
\frac{{\bf{k}}\times \frac{\rm d}{{\rm
d}t}{\bf{k}}}{k^{2}}={\bf{B}}_{\rm g}, \eqnum{28} \label{eq27}
\end{equation}
where the gravitomagnetic field strength ${\bf{B}}_{\rm g}$ is
determined by the field equations of gravitation such as Eq.
(\ref{eq13}) and Einstein$^{,}$s equation of general relativity.
It follows from the geodesic equation in the weak-field
approximation that the acceleration due to gravitational Lorentz
force is
\begin{equation}
\frac{1}{k}\frac{\rm d}{{\rm d}t}{\bf{k}}=-\frac{{\bf{k}}}{k}\times \left({\nabla \times \vec{g%
}}\right)  \eqnum{29}  \label{eq28}
\end{equation}
with $\frac{{\bf{k}}}{k}$ being the velocity vector of the photon,
and the gravitomagnetic field strength is ${\bf{B}}_{\rm g}=\nabla
\times {\bf {g}},$ where
${\bf {g}}=(g^{01},g^{02},g^{03}).$ Substitution of Eq. (\ref{eq27}) into Eq. (%
\ref{eq28}) yields
\begin{equation}
\frac{1}{k}\frac{\rm d}{{\rm d}t}{\bf{k}}=-\frac{{\bf{k}}}{k}\times \left( \frac{{\bf{k}}%
\times \frac{\rm d}{{\rm d}t}{\bf{k}}}{k^{2}}\right) .  \eqnum{30}
\label{eq29}
\end{equation}
Since
\begin{equation}
{\bf{k}}\cdot {\bf{k}}=k^{2},\quad {\bf{k}}\cdot \frac{\rm d}{{\rm
d}t}{\bf{k}}=0, \eqnum{31}  \label{eq30}
\end{equation}
Eq. (\ref{eq29}) is proved to be an identity. This, therefore, implies that Eq. (%
\ref{eq27}) is truly the equation of motion of a photon in
gravitomagnetic fields.

For the time-dependent gravitomagnetic fields, similar to the case
of the photon propagating inside the noncoplanarly curved optical
fiber that is wound smoothly on a large enough
diameter\cite{Chiao,Tomita,Kwiat}, the photon propagating in the
gravitomagnetic field would also give rise to a geometric phase,
which can be calculated by making use of the Lewis-Riesenfeld
invariant theory\cite{Lewis} and the invariant-related unitary
transformation formulation\cite{Gao}, and the result is
\begin{equation}
\phi _{\pm}^{\rm (g)}=\pm \textstyle\int
_{0}^{t}\dot{\gamma}(t^{^{\prime }})[1-\cos \lambda (t^{^{\prime
}})]{\rm d}t^{^{\prime }}  \eqnum{32} \label{eq100}
\end{equation}
with $\pm $ corresponding to the eigenvalue $\sigma =\pm 1$ of the
helicity $\frac{{\bf{k}}(t)}{k}\cdot {\bf{J}}$ of the photon. The
time-dependent
parameters, $\gamma $ and $\lambda $, are so defined that $\frac{{\bf{k}}(t)}{k%
}=\left(\sin \lambda (t)\cos \gamma (t),\sin \lambda (t)\sin
\gamma (t),\cos \lambda (t)\right).$ Differing from the dynamical
phase that is related to the energy, frequency or velocity of a
particle or a quantum system, geometric phase is dependent only on
the geometric nature of the pathway along which the system
evolves. For the case of adiabatic process where $\lambda $ does
not explicitly involve time $t$, Eq. (\ref{eq100}) is reduced to
\begin{equation}
\Delta \theta =\pm 2\pi (1-\cos \lambda )  \eqnum{33}  \label{eq331}
\end{equation}
in one cycle over the parameter space of the helicity $\frac{{\bf{k}}(t)}{k}%
\cdot {\bf{J}}$. It follows from Eq. (\ref{eq16}) that $2\pi
(1-\cos \lambda ) $ is the expression for a solid angle in
${\bf{k}}(t)$ space, which presents the geometric properties of
time evolution of the interaction between the gravitomagnetic
field and photon spin ( gravitomagnetic moment ).

(2) Taking the effects of gravitomagnetic fields and
gravitomagnetic charges into consideration is believed to be of
essential significance in resolving some problems and paradoxes.
An illustrative example that would be briefly discussed in what
follows may be regarded as an application of gravitomagnetic
fields ( matter ) to the cosmological constant problem. The
gravitational analogue of Meissner effect in superconductivity is
gravitational Meissner effect. Due to the conservation of
momentum, mass-current density is conserved before and after the
scatterings of particles in perfect fluid, which is analogous to
the superconductivity of superconducting electrons in
superconductors cooled below $T_{c}$. Since gravitational field
equation of linear approximation is similar to the Maxwell's
equation in electrodynamics, one can predict that gravitational
Meissner effect arises also in perfect fluid. The author think
that the investigation of both the effect of gravitomagnetic
matter and gravitational Meissner effect may provide us with a
valuable insight into the problem of cosmological constant and
vacuum gravity\cite{Weinberg,Datta,Alvarenga,Cap}. Given that the
vacuum matter is perfect fluid, the gravitoelectric field (
Newtonian gravitational field ) produced by the gravitoelectric
charge ( mass ) of the vacuum quantum fluctuations is exactly
cancelled by the gravitoelectric field due to the induced current
of the gravitomagnetic charge of the vacuum quantum fluctuations;
the gravitomagnetic field produced by the gravitomagnetic charge (
dual mass ) of the vacuum quantum fluctuations is exactly
cancelled by the gravitomagnetic field due to the induced current
of the gravitoelectric charge ( mass current ) of the vacuum
quantum fluctuations. Thus, at least in the framework of
weak-field approximation, the extreme space-time curvature of
vacuum caused by its large energy density does not arise, and the
gravitational effects of cosmological constant\footnote{For the
more present considerations regarding the cosmological constant
problem, see, for example, hep-ph/0002297 ( by E. Witten ) and
astro-ph/0005265 ( by S. Weinberg ).} is eliminated by the
contributions of the gravitomagnetic charge ( dual mass ).

Additionally, in 1998, Anderson {\it et al.} reported that, by
ruling out a number of potential causes, radio metric data from
the Pioneer 10/11, Galileo and Ulysses spacecraft indicate an
apparent anomalous, constant, acceleration
acting on the spacecraft with a magnitude $\sim 8.5\times $ $10^{-8}$cm/s$%
^{2}$ directed towards the Sun\cite{Anderson}. Is it the effects
of dark matter or a modification of gravity? Unfortunately,
neither easily works. By taking the cosmic mass, $M=10^{53}$ kg,
and cosmic scale, $R=10^{26}$ m, calculation shows that this
acceleration is just equal to the value of field strength on the
cosmic boundary due to the total cosmic mass. This fact leads us
to consider a theoretical mechanism to interpret this anomalous
phenomenon. The author favors that the gravitational Meissner
effect may serve as a possible interpretation. Here we give a
rough analysis, which contains only the most important features
rather than the precise details of this theoretical explanation.
Parallel to London$^{,}$s electrodynamics of superconductivity, it
shows that gravitational field may give rise to an {\it effective
rest mass }$m_{g}=\frac{\hbar }{c^{2}}\sqrt{8\pi G\rho _{m}}$ due
to the self-induced charge current \cite{Hou}, where $\rho _{m}$
is the mass
density of the universe. Then one can obtain that $\frac{\hbar }{m_{g}c}%
\simeq 10^{26}$m that approximately equals $R$, where the mass
density of the universe is taken to be $\rho _{m}=0.3\rho
_{c}$\cite{Pea} with $\rho _{c}\simeq 2\times 10^{-26}$kg/m$^{3}$
being the critical mass density. An added constant acceleration,
$a$, may result from the Yukawa potential and can be written
as\footnote{It may also be calculated as follows: $a =\frac{GM}{2}\left( \frac{m_{g}c}{\hbar }\right) ^{2}=\frac{GM}{c^{2}}%
(4\pi \rho _{m}G)=\frac{GM}{c^{2}R}\frac{G(4\pi R^{3}\rho _{m})}{R^{2}}\simeq \frac{GM}{%
R^{2}}$, where use is made of $\frac{GM}{c^{2}R}\simeq {\mathcal
O}(1),4\pi R^{3}\rho _{m}\simeq M$, which holds when the
approximate estimation is performed.}

\begin{equation}
a=\frac{GM}{2}\left( \frac{m_{g}c}{\hbar }\right) ^{2}\simeq \frac{GM}{R^{2}}%
.  \eqnum{34}  \label{eq31}
\end{equation}
Note, however, that it is an acceleration of repulsive force directed,
roughly speaking, from the center of the universe. By analyzing the NASA$%
^{,} $s Viking ranging data, Anderson, Laing, Lau {\it et al.}
concluded that the anomalous acceleration does not act on the body
of large mass such as the Earth and Mars. If gravitational
Meissner effect only affected the gravitating body of large mass
or large scale rather than spacecraft ( perhaps the reason lies in
that small-mass flow cannot serve as self-induced charge current,
which deserves to be further investigated ), then seen from the
Sun or Earth, there exists an added attractive force acting on the
spacecraft. This added force give rise to an anomalous, constant,
acceleration directed towards the Sun or Earth. It should be
emphasized that the above theoretical theme is only a potential
interpretation of this anomalous gravitational phenomenon.
However, the sole reason that the above resolution of the
anomalous acceleration is somewhat satisfactory lies in that no
adjustable parameters exist in this theoretical framework. It is
one of the most important advantages in the above mechanism of
gravitational Meissner effect, compared with some possible
theories of modification of gravity\cite{Nieto}, which are always
involving several parameters that cannot be determined by theory
itself. These theories of modification of gravity were applied to
the problem of the anomalous acceleration but could not calculate
the value of the anomalous acceleration.

\section{CONCLUDING REMARKS}

In summary, in the present paper the author investigates some
geometric effects, gravitational analogues of electromagnetic
phenomena, and the field equation of gravitomagnetic matter ( dual
matter ) as well as its static spherically symmetric solution.
Differing from the symmetric property ( with respect to the
indices of tensors ) of gravitational field equation of
gravitoelectric matter, the field equation of gravitomagnetic
matter possesses the antisymmetric property. This, therefore,
implies that the number of the non-analytic metric functions is no
more than 6. Although we have no observational evidences for the
existence of gravitomagnetic charge, it is still of essential
significance to investigate the gravity theory of the topological
dual mass\footnote{Many investigations concerning the
gravitomagnetic monopole and its quantization as well as the
related effects and phenomena such as the gravitational lensing
and the orbits of motion of matter in NUT space were performed.
Readers can be referred to [Rev. Mod. Phys. {\bf 70}(2),
427-445(1998)]( see also in gr-qc/9612049 ) and the references
therein. However, we think in these references the gravitational
field equation of gravitomagnetic monopole may get less attentions
than it deserves.}.

Some physically interesting problems associated with
gravitomagnetic fields are proposed, of which the most interesting
investigation is the potential solution to the anomalous
acceleration acting on Pioneer spacecrafts by means of the
mechanism of gravitational Meissner effect. The theoretical
resolution of this problem is not very definite at present, for it
cannot account for why the gravitational Meissner effect does not
explicitly affect the body of small mass. This curiosity deserves
further considerations.

Since with foreseeable improvements in detecting and measuring
technology, it is possible for us to investigate quantum mechanics
in weak-gravitational fields, the above effects and phenomena
deserve further detailed investigations. Work in this field is
under consideration and will be published elsewhere.
\\ \\
\\ \\
{\bf APPENDIX}
\\ \\
It is necessary to analyze the relation between the coupling
parameters, $\kappa _{1}$ and $\kappa _{2}$, under the low-motion
and weak-field approximation. Substitution of the expression
(\ref{eq16}) for $S^{\mu \nu }$ into Eq. (\ref{eq14}), where
\begin{equation}
S_{\quad ;\nu }^{\mu \nu }=\frac{\partial S^{\mu \nu }}{\partial x^{\nu }}+%
\frac{1}{2}S^{\mu \lambda }g^{\sigma \tau }\frac{\partial g_{\sigma \tau }}{%
\partial x^{\lambda }},  \eqnum{A1}  \label{A1}
\end{equation}
yields
\begin{eqnarray}
\kappa _{1}g^{0}\frac{\partial }{\partial x^{0}}{\bf{v}}
&=&2\kappa
_{2}[\nabla \times {\bf {g}}-{\bf{v}}\times (\nabla g^{0}-\frac{\partial }{%
\partial x^{0}}{\bf {g}})]-2\kappa _{2}{\bf {g}}\times (\frac{\partial }{%
\partial x^{0}}{\bf{v}}+\nabla g^{0})  \nonumber \\
&&+\kappa _{1}g^{0}\frac{\partial }{\partial x^{0}}{\bf
{g}}-2\kappa _{2}g^{0}(\nabla g^{0}\times {\bf{v}})  \eqnum{A2}
\label{A2}
\end{eqnarray}
with ${\bf{v}}$ being the velocity of the tested gravitomagnetic
monopole. It
is apparent that $\nabla \times {\bf {g}}-{\bf{v}}\times (\nabla g^{0}-\frac{%
\partial }{\partial x^{0}}{\bf {g}})$ is the expression associated with
gravitational Lorentz force density. Note that in Eq. (\ref{A2}), $\kappa
_{1},\kappa _{2}$ are considered coupling constants. However, further
analysis shows that at least one of them is not a constant, and if the
relation
\begin{equation}
\kappa _{1}g^{0}=2\kappa _{2}  \eqnum{A3}  \label{A3}
\end{equation}
between them is assumed, then Eq. (\ref{A2}) may be rewritten as
\begin{equation}
\frac{\partial }{\partial x^{0}}{\bf{v}}=[\nabla \times {\bf
{g}}-{\bf{v}}\times (\nabla g^{0}-\frac{\partial }{\partial
x^{0}}{\bf {g}})],  \eqnum{A4} \label{A4}
\end{equation}
where we temporarily ignore the second and third terms, which can
be considered the small terms, on the right-hand side of Eq.
(\ref{A2}) and the derivative term of coupling coefficients with
respect to space-time coordinates $x^{\mu }$. It is well known
that the form of Eq. (\ref{A4}) is the equation of motion of a
particle acted upon by the Lorentz force. Hence, Eq. (\ref{eq14})
is believed to be the generally relativistic equation of motion of
gravitomagnetic monopole in the Riemann space-time.

\end{document}